\documentclass{aa} 

\usepackage{graphicx} 

\begin{document}
   
\title{Oxygen abundances in dwarf irregular galaxies and the 
          metallicity -- luminosity relationship}

\author{ L.S. Pilyugin }

   \institute{   Main Astronomical Observatory
                 of National Academy of Sciences of Ukraine,
                 Goloseevo, 03680 Kiev-127, Ukraine, \\
                 (pilyugin@mao.kiev.ua)
                 }

\offprints{L.S. Pilyugin }

\date{Received 20 December 2000 / accepted 27 April 2001}

\abstract{
The low-luminosity dwarf irregular galaxies are considered.
The oxygen abundances in H\,{\sc ii} regions of dwarf irregular galaxies  were 
recalculated from published spectra through the recently suggested P --
method. It has been found  that the metallicity of low-luminosity dwarf 
irregular galaxies, with a few exceptions, correlates well with galaxy 
luminosity. The dispersion of oxygen abundances around the metallicity -- 
luminosity relationship increases with decreasing of galaxy luminosity, 
as was found by Richer \& McCall (1995). No relationship between the 
oxygen abundance and the absolute magnitude in the blue band for irregular
galaxies obtained by Hidalgo-G\'amez \& Olofsson (1998) can be explained by the
large uncertainties in the oxygen abundances derived through the $T_{e}$ --
method, that in turn can be explained by the large uncertainties in the 
measurements of the strengths of the weak oxygen line $[OIII] \lambda 4363$ 
used in the $T_{e}$ -- method.
}

\titlerunning{Oxygen abundances in dwarf irregular galaxies and the 
          metallicity -- luminosity relationship}

\authorrunning{L.S.~Pilyugin}  

\maketitle

\section{Introduction}

Twenty years ago Lequeux et al. (1979) revealed that the oxygen abundance 
correlates with total galaxy mass for irregular galaxies, in the sense that
the higher the total mass, the higher the heavy element content. Since the 
galaxy mass is a poorly known parameter, the metallicity -- luminosity relation 
instead of the mass -- metallicity relation is usually considered (Skillman et 
al. 1989a, Richer \& McCall 1995, Hidalgo-G\'amez \& Olofsson 1998, Hunter \& 
Hoffman 2000, Pilyugin \& Ferrini 2000b, among others). It has been found that 
the characteristic gas-phase abundances (the oxygen abundance at a predetermined 
galactocentric distance) and luminosities of spiral galaxies are also correlated 
(Garnett \& Shields 1987, Vila-Costas \& Edmunds 1992, Zaritsky et al. 1994, 
Garnett et al. 1997), and this relationship maps almost directly on to the 
metallicity -- luminosity relationship of irregular galaxies (Zaritsky et al. 
1994, Garnett et al. 1997). 

Richer \& McCall (1995) have revealed a prominent feature of their
metallicity -- luminosity relation for irregular galaxies: they have found more 
scatter at low luminosities, though they found less at high luminosities. The 
onset of this scatter seems to occur at $M_{B} \sim - 15$ or $\log L_{B} \sim 8.2$. 
Moreover, Hidalgo-G\'amez \& Olofsson (1998) have found that there is no 
relationship between the oxygen abundance and the absolute magnitude in the blue 
band for dwarf irregular galaxies ($M_{B} > - 17$ or $\log L_{B} < 9$). Hunter \& 
Hoffman (2000) have found that their samples do generally cluster around the 
relationship between $M_{B}$ and O/H derived by Richer \& McCall (1995) but do 
not themselves define a linear relationship very well, appearing more as a cloud 
of points with a large scatter around the line. In particular, the scatter 
becomes larger for $M_{B}$ $>$ --16. Hunter \& Hoffman (2000) have concluded 
that the relationship between O/H and $M_{B}$ is very general over the Hubble 
sequence but the scatter is very large and discerning the trend over a limited 
parameter range is hard. They noted that perhaps the interesting science is in 
this scatter, although a part of this scatter is undoubtedly due to the 
uncertainties in determining O/H and $M_{B}$. 

The most precise method of determining the abundances of H\,{\sc ii} regions 
requires the detection of $[OIII] \lambda 4363$ emission line (the $T_{e}$ -- 
method). The $[OIII] \lambda 4363$ emission line appears in high excitation 
spectra of oxygen-poor H\,{\sc ii} regions and is usually undetectable in 
spectra of oxygen-rich H\,{\sc ii} regions. Then, in the general case, the 
precision of the oxygen abundance determination in oxygen-poor H\,{\sc ii} 
regions is higher than in oxygen-rich H\,{\sc ii} regions. The derived oxygen 
abundances of metal-poor H\,{\sc ii} regions in blue compact galaxies possessing 
very bright emission lines are accurate to within 0.05 dex (Izotov \& Thuan 
1999). However, many irregular galaxies have no bright H\,{\sc ii} regions with 
readily measured emission lines. As was noted by Hidalgo-G\'amez \& Olofsson 
(1998) the uncertainties in the intensity of the line $[OIII] \lambda 4363$ in 
spectra of H\,{\sc ii} regions in dwarf irregular galaxies reported in the 
literature fluctuate between 11$\%$ and 120$\%$. Thus, in reality, the precision 
of oxygen abundance determination in dwarf irregular galaxies seems to be 
rather low.

In our recent work (Pilyugin 2000 (Paper I), 2001 (Paper II)) a new method for 
oxygen abundance determination in H\,{\sc ii} regions (the P -- method) has been 
constructed, starting from the idea of McGaugh (1991) that the strong oxygen 
lines ($[OII] \lambda \lambda 3727, 3729$ and $[OIII] \lambda \lambda 4959, 5007$) 
contain the necessary information for determination of accurate abundances in 
H\,{\sc ii} regions. By comparing  oxygen abundances in bright H\,{\sc ii} 
regions derived (with high precision) through the $T_{e}$ -- method 
O/H$_{T_{e}}$ and those derived through the suggested P -- method O/H$_{P}$ it 
has been found that the precision of oxygen abundance determination with the 
P -- method is comparable to that  of the $T_{e}$ -- method. Then it can be 
expected that in faint H\,{\sc ii} regions, in which the temperature-sensitive 
line $[OIII] \lambda 4363$ is measured with large uncertainty, the P -- method 
provides more realistic oxygen abundances than the T$_{e}$ -- method since 
only the strong (and as a consequence, more readily measurable) oxygen lines 
are used in the P -- method. Moreover, the P -- method is workable in the 
cases when the temperature-sensitive line $[OIII] \lambda 4363$ is undetectable.
Thus, we can expect that the application of the P -- method to the oxygen 
abundance determination in faint H\,{\sc ii} regions of dwarf irregular galaxies 
allows us to refine the oxygen abundances in H\,{\sc ii} regions with 
low-precision measurements of line $[OIII] \lambda 4363$ and to determine the 
realistic oxygen abundances in H\,{\sc ii} regions with undetectable line 
$[OIII] \lambda 4363$. We hope that it can clarify whether the luminosity -- 
metallicity relationship for irregular galaxies still persists or disappears 
at the low-luminosity end. This is a goal of the present study.

\section{The Z-L relationship of low-luminosity irregular galaxies}

\subsection{The preliminary remarks}

The relation of the type O/H=f(P,$R_{3}$) between oxygen abundance and the 
values of P and R$_{3}$ has been derived empirically in Papers I and II using 
the available oxygen abundances determined via measurement of the 
temperature-sensitive line ratio [OIII]4959,5007/[OIII]4363. Notations similar 
to those in Papers I and II will be adopted here:
$R_{2}$ = $I_{[OII] \lambda 3727+ \lambda 3729} /I_{H\beta }$, 
$R_{3}$ = $I_{[OIII] \lambda 4959+ \lambda 5007} /I_{H\beta }$, 
$R$ = $I_{[OIII] \lambda 4363} /I_{H\beta }$, 
$R_{23}$ =$R_{2}$ + $R_{3}$, $X_{23}$ = log$R_{23}$, 
and P = $R_{3}$/$R_{23}$. The excitation index P used in Paper II and indexes
$p_{2}$ and $p_{3}$ used in Paper I are related through simple expressions:
$p_{3}$ = logP and  $p_{2}$ = log(1-P). The following equations for the oxygen 
abundance determination in low-metallcity H\,{\sc ii} regions have been 
suggested in Paper I
\begin{equation}
X^{*}_{3} = X_{3}^{obs} - 2.20 \; p_{3},
\label{equation:xs-p}
\end{equation}
and
\begin{equation}
12 + \log(O/H)_{P}  = 6.35 + 1.45 \; X^{*}_{3} .
\label{equation:oh-xs}
\end{equation}
Eq.(\ref{equation:xs-p}) and Eq.(\ref{equation:oh-xs}) can be rewrtitten as
\begin{equation}
12 + \log (O/H)_{P}  = 6.35 + 1.45 \; \log R_{3} - 3.19\; \log P.
\label{equation:oh-p}
\end{equation}
Eq.(\ref{equation:oh-p}) shows that the positions in the R$_{3}$ -- P 
diagram can be calibrated in terms of oxygen abundance (Fig.\ref{figure:r3-p}).

The relationship between oxygen abundance and strong line intensities is double 
valued with two distinctive parts named usually as the lower and upper branches 
of the R$_{23}$ -- O/H diagram, and so one has to know in advance on which 
branch of the R$_{23}$ -- O/H diagram the H\,{\sc ii} region lies. The above 
expression for the oxygen abundance determination in H\,{\sc ii} regions, 
Eq.(\ref{equation:oh-p}), is valid for H\,{\sc ii} regions with 12+log(O/H) less 
than around 8 (Paper I). According to the metallicity -- luminosity relationship 
for irregular galaxies after Richer \& McCall (1995), the oxygen abundances in 
the low-luminosity irregular galaxies ($M_{B}$ $>$ --15$\div$--16) are expected 
to lie in this range. Then it has been adopted here that the H\,{\sc ii} regions 
in the low-luminosity irregular galaxies ($M_{B}$ $>$ --15$\div$--16) lie on 
the lower branch of the R$_{23}$ -- O/H diagram. Furthermore, only the 
high-excition (P $>$ 0.5) H\,{\sc ii} regions will be considered here because 
Eq.(\ref{equation:xs-p}) and Eq.(\ref{equation:oh-xs}) (linear approximation) 
have been derived based on the high-excition H\,{\sc ii} regions.

\subsection{Hidalgo-G\'amez \& Olofsson sample}

\begin{figure}
\resizebox{1.00\hsize}{!}{\includegraphics[angle=0]{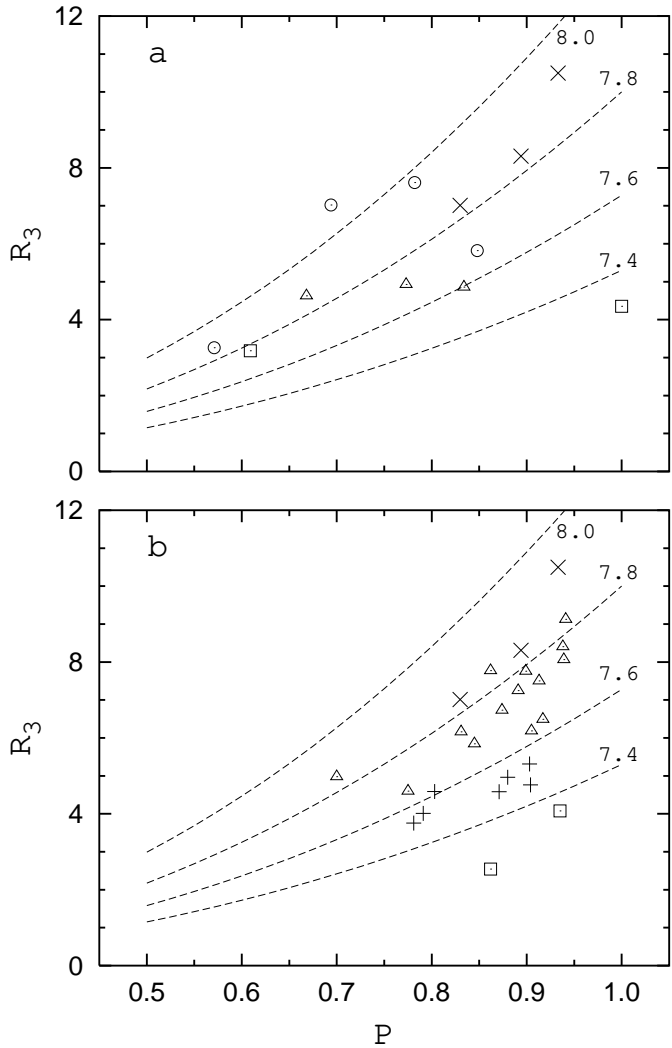}}\caption{
The R$_{3}$ -- P diagram for H\,{\sc ii} regions with M$_{B}$ $>$ --16 from 
Hidalgo-G\'amez \& Oloffson (1998) sample (panel {\it a}) and for "calibrating 
H\,{\sc ii} regions" from Paper I (panel {\it b}). The squares are H\,{\sc ii} 
regions with 12+logO/H $<$ 7.4, the pluses are those with 7.4 $\leq$ 12+logO/H 
$<$ 7.6, the triangles are those with 7.6 $\leq$ 12+logO/H $<$ 7.8, the crosses 
are those with 7.8 $\leq$ 12+logO/H $<$ 8.0, and the circles are those with  
12+logO/H $\geq$ 8.0. The dashed curves are R$_{3}$ -- P relations predicted 
by the calibration for fixed values of O/H. Each curve is labeled with 
the corresponding value of 12+logO/H.}
\label{figure:r3-p}
\end{figure}

Firstly the Hidalgo-G\'amez \& Olofsson (1998) sample of irregular galaxies will 
be considered. The R$_{3}$ -- P diagram for H\,{\sc ii} regions with M$_{B}$ 
$>$ --16 from Hidalgo-G\'amez \& Oloffson (1998) sample is shown in the
Fig.\ref{figure:r3-p} (panel {\it a}). The squares are H\,{\sc ii} regions with 
12+logO/H $<$ 7.4, the pluses are those with 7.4 $\leq$ 12+logO/H $<$ 7.6, the 
triangles are those with 7.6 $\leq$ 12+logO/H $<$ 7.8, the crosses are those 
with 7.8 $\leq$ 12+logO/H $<$ 8.0, and the circles are those with 12+logO/H 
$\geq$ 8.0. The dashed curves are the R$_{3}$ -- P relations obtained from 
Eq.(\ref{equation:oh-p}) for fixed values of O/H. Each curve is labeled with the 
corresponding value of 12+logO/H.  For comparison the R$_{3}$ -- P diagram for 
"calibrating H\,{\sc ii} regions" from Paper I is also shown in 
Fig.\ref{figure:r3-p} (panel {\it b}). Examination of Fig.\ref{figure:r3-p} 
(panel {\it a}) shows that the oxygen abundances O/H$_{T_{e}}$ derived through 
the classical T$_{e}$ -- method are in conflict with the oxygen abundances 
corresponding to their positions in the R$_{3}$ -- P for a number of H\,{\sc ii} 
regions from the Hidalgo-G\'amez \& Olofsson (1998) sample of irregular galaxies. 
It can be explained by the large uncertainties in the oxygen abundances derived 
through the $T_{e}$ -- method, that in turn can be explained by the large 
uncertainties in the measurements of the strengths of the weak oxygen line 
$[OIII] \lambda 4363$ used in the $T_{e}$ -- method.

\begin{figure}
\resizebox{1.00\hsize}{!}{\includegraphics[angle=0]{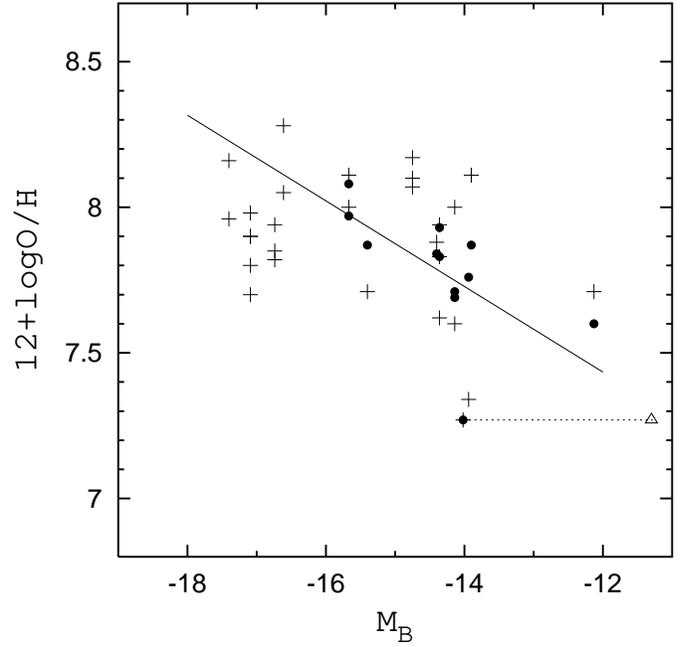}}
\caption{The metallicity -- luminosity diagram for dwarf irregular galaxies. The 
metallicity -- luminosity diagram after Hidalgo-G\'amez \& Olofsson (1998) based 
on the oxygen abundances determined with the $T_{e}$ -- method is presented by 
the pluses. The metallicity -- luminosity diagram based on the oxygen abundances 
derived with the P -- method is shown by the filled circles. The positions of 
Leo A with luminosity from Hidalgo-G\'amez \& Olofsson and with luminosity adopted 
here (triangle) are connected with dashed line. The solid line is the L -- Z 
relationship after Richer \& McCall (1995).}
\label{figure:hidalgo}
\end{figure}

The oxygen abundances in H\,{\sc ii} regions of low-luminosity ($M_{B}$ $>$ 
--16) irregular galaxies have been recalculated through the P -- method. The 
intensities of $[OII] \lambda \lambda 3727, 3729$ and 
$[OIII] \lambda \lambda 4959, 5007$ lines were taken from the sources cited by 
Hidalgo-G\'amez \& Olofsson (1998): Skillman, Kennicut and Hodge 1989 
(DDO 47, Leo A, Sext A); Moles, Aparacio and Masegosa 1990 (Sext B, GR 8);
Gonz\'alez-Riestra, Rego and Zamorano 1988 (Mkn 178); Heydari-Malayeri, Melnick 
and Martin 1990 (IC 4662); Webster, Longmore, Hawarden and Mebold 1983 (IC 5152); 
Hodge \& Miller 1995 (WLM). The NGC6822 is excluded from consideration here, 
this galaxy will be discussed below. The oxygen abundances have been recalculated 
with the P -- method in 12 H\,{\sc ii} regions of 9 dwarf irregular galaxies. 
The H\,{\sc ii} regions A1 and A2 in IC4662 have oxygen abundances 12+logO/H$_{P}$ 
= 7.97 and 8.08, respectively. These H\,{\sc ii} regions seems to belong to the 
transition zone of the R$_{23}$ -- O/H diagram, and in the strict sense 
Eq.(\ref{equation:oh-p}) cannot be used for oxygen abundance determination in 
these H\,{\sc ii} regions since Eq.(\ref{equation:oh-p}) was derived for 
H\,{\sc ii} regions with 12+logO/H $\leq$ 7.95 (the lower branch of the R$_{23}$ 
-- O/H diagram). The metallicity -- luminosity diagram for dwarf irregular 
galaxies from Hidalgo-G\'amez \& Olofsson (1998) is represented in 
Fig.\ref{figure:hidalgo} by the pluses. The metallicity -- luminosity diagram 
for dwarf irregular galaxies with M$_{B}$ $>$ -- 16 with the oxygen abundances 
recalculated through Eq.(\ref{equation:oh-p}) is shown in Fig.\ref{figure:hidalgo} 
by the filled circles. In order to clearly recognize the influence of the oxygen 
abundance redetermination on the metallicity -- luminosity diagram, the same 
values of $M_{B}$ as in Hidalgo-G\'amez \& Olofsson were used.

It can be easily seen in Fig.\ref{figure:hidalgo} that the metallicity -- 
luminosity relation for dwarf irregular galaxies based on the oxygen abundances 
derived with the P -- method shows appreciably less scatter than that based on 
the oxygen abundances derived with the $T_{e}$ -- method. It is strong evidence 
in favor of the large scatter in the low-luminosity end of the metallicity -- 
luminosity relation for irregular galaxies is explained mainly by the 
uncertainty in oxygen abundance determination with the $T_{e}$ -- method.

It should be noted that the abundances used by Hidalgo-G\'amez \& Olofsson were 
not taken directly from the sources listed but have been recalculated in a 
consistent way and as a consequence the differences in techniques and atomic 
data do not contribute to the scatter. Then, the increased scatter in oxygen 
abundances derived with the $T_{e}$ -- method, compared to the scatter in 
oxygen abundances derived with the P -- method seems to be explained by the 
observational uncertainties in the $[OIII] \lambda 4363$ line strengths.

A well defined trend is seen between $M_{B}$ and O/H$_{P}$, 
Fig.\ref{figure:hidalgo}. The deviations of O/H$_{P}$  from a linear 
relationship for all, with one exception, of the H\,{\sc ii} regions are within 
$\pm$ 0.1dex. The point with a large deviation (in excess of 0.4 dex) corresponds 
to the dwarf galaxy Leo A. The H\,{\sc ii} region observed in Leo A is a 
planetary nebula with no detectable $[OII] \lambda \lambda 3727, 3729$ lines; 
P = 1 was adopted in the oxygen abundance determination through the P -- method. 
It cannot be excluded that the P -- method calibrated on the basis of 
H\,{\sc ii} regions is inapplicable in the case of planetary nebulae. It should 
be noted, however, that the oxygen abundance in Leo A derived with the P -- 
method agrees remarkably with the oxygen abundance derived by Skilman, Kennicutt 
and Hodge (1989). The deviation of Leo A from the relationship seems to be 
caused by the uncertainty in the integrated B-band absolute magnitude adopted by 
Hidalgo-G\'amez \& Olofsson rather than the uncertainty in the oxygen abundance. 
The position of Leo A with a new integrated B-band absolute magnitude (see next 
section) is shown in Fig.\ref{figure:hidalgo} by triangles (the positions of 
Leo A in the Z -- L diagram with a new absolute magnitude and with absolute 
magnitude adopted by Hidalgo-G\'amez \& Olofsson are connected with a dashed line). 

Thus, the determination of oxygen abundances in H\,{\sc ii} regions of 
low-luminosity irregular galaxies from the Hidalgo-G\'amez \& Olofsson sample with 
the P -- method suggests that there is a well defined trend between $M_{B}$ and 
O/H$_{P}$, and the large scatter in the low-luminosity end of the metallicity -- 
luminosity relation obtained by Hidalgo-G\'amez \& Olofsson (1998) is explained
mainly by the uncertainty in oxygen abundance determination with the $T_{e}$ --
method. In order to verify this conclusion, a larger sample of low-luminosity 
irregular galaxies will be considered in the next section.

\subsection{The Z-L relationship of low-luminosity irregular galaxies}


\begin{table*}
\caption[]{\label{table:OH}
Characteristics of the galaxies in the present sample}
\begin{center}
\begin{tabular}{lllllllcc} \hline \hline
            &            &            &              &           &                          &             &                     &                     \\ 
galaxy      & other name & H\,{\sc ii} region &  $M_{B}$     & reference &       O/H$^{1}$          & reference   & (O/H)$_{P}^{1}$     &  (O/H)$_{P}^{1}$    \\  
            &            & or spectrum&              &           &                          &             & individual          &   average           \\ 
            &            & label      &              &           &                          &             & H\,{\sc ii} region          &   for galaxy        \\  \hline
Sextans B   & DDO 70     &            & -13.8        &    M      &  8.11                    &    MAM      &{\it 7.87}           &{\bf 7.86}           \\
            &            &   No 2     &              &           &7.56$^{*}$ (7.54 -- 7.57) &    SKH      &{\it 7.85}           &                     \\
Sextans A   & DDO 75     &   No 1     & -14.2        &    M      & 7.40 (6.70 -- 7.65)      &    SKH      &{\it 7.76}           &{\bf 7.71}           \\
            &            &   No 1h    &              &           & 7.58 (6.90 -- 7.83)      &    SKH      &{\it 7.65}           &                     \\
GR 8        & DDO 155    &    H2a     & -11.2        &    M      & 7.66                     &    MAM      &{\it 7.61}           &{\bf 7.60}           \\
            &            &    H2b     &              &           &  7.71                    &    MAM      &{\it 7.60}           &                     \\
WLM         & DDO 221    &   No 7     & -13.9        &    M      &  7.72 (7.64 -- 7.78)     &    HM       &{\it 7.71}           &{\bf 7.78}           \\
            &            &   No 9     &              &           &  7.81 (7.73 -- 7.88)     &    HM       &{\it 7.69}           &                     \\
            &            &   No 1     &              &           &  7.78 (7.48 -- 7.95)     &    STM      &{\it 7.79}           &                     \\
            &            &   No 2     &              &           &  7.70 (7.30 -- 7.90)     &    STM      &{\it 7.89}           &                     \\
UGC 4483    &            &   INT      & -12.8        &   STKGT   &  7.50 (7.46 -- 7.54)     &    STKGT    &{\it 7.45}           &{\bf 7.47}           \\
            &            &   McD      &              &           &  7.48 (7.42 -- 7.54)     &    STKGT    &{\it 7.46}           &                     \\
            &            &   WHT      &              &           &  7.57 (7.51 -- 7.62)     &    STKGT    &{\it 7.50}           &                     \\
Mkn 178     &            &   A        & -14.36       &    H-GO   &  7.72                    &    G-RRZ    &{\it 7.83}           &{\bf 7.88}           \\
            &            &   B        &              &           &  7.73                    &    G-RRZ    &{\it 7.92}           &                     \\
M81dB       & UGC 5423   &            & -12.9        &    MH94   &  7.98 (7.70 -- 8.15)     &    MH       &{\it 7.81}           &{\bf 7.81}           \\
UGC 6456    &            &   Case 1   & -13.24       &    RM     &  7.77                    &    TBDS     &{\it 7.68}           &{\bf 7.71}           \\
            &            &   Case 2   &              &           &  7.76                    &    TBDS     &{\it 7.75}           &                     \\
Leo A       & DDO 69     &            & -11.3        &    M      &  7.28 (7.08 --7.42)      &    SKH      &{\it 7.27}           &{\bf 7.27}           \\
DDO 187     &            &            & -13.4        &    SKH    &7.36$^{*}$ (7.23 -- 7.46) &    SKH      &{\it 7.62}           &{\bf 7.62}           \\
DDO 47      &            &  No 1      & -14.4        &    H-GO   &  7.89 (7.64 -- 8.05)     &    SKH      &{\it 7.84}           &{\bf 7.86}           \\
            &            &  No 3      &              &           &7.71$^{*}$ (7.66 -- 7.75) &    SKH      &{\it 7.87}           &                     \\
UGCA 292    &            &  No 1      & -11.43       &    vZ     &  7.28 (7.23 -- 7.33)     &    vZ       &{\it 7.26}           &{\bf 7.22}           \\
            &            &  No 2      &              &           &  7.32 (7.26 -- 7.38)     &    vZ       &{\it 7.17}           &                     \\
DDO 167     &            &            & -13.3        &    SKH    &  7.66 (7.23 -- 7.88)     &    SKH      &{\it 7.81}           &{\bf 7.81}           \\  
SagDIG      &            &            & -12.1        &    M      &  7.36$^{*}$              &    STM      &{\it 7.48}           &{\bf 7.48}           \\
A1116+51    &            &            & -14.99$^{2}$ &    KD     &  7.55 (7.35 -- 7.75)     &    KD       &{\it 7.76}           &{\bf 7.76}           \\  
A1228+12    &            &            & -14.57$^{2}$ &    KD     &  7.64 (7.57 -- 7.71)     &    KD       &{\it 7.79}           &{\bf 7.79}           \\  
A2228-00    &            &            & -14.85$^{2}$ &    KD     &  7.62 (7.56 -- 7.68)     &    KD       &{\it 7.81}           &{\bf 7.81}           \\  
ESO 245-G05 &            &  19        & -15.5        &  H-G99    &  7.80 (7.79 -- 7.81)     & H-G99       &{\it 7.98}           &{\bf 7.94}           \\  
            &            &  12        &              &  H-G99    &  7.59 (7.57 -- 7.61)     & H-G99       &{\it 7.89}           &                     \\ 
DDO 53      &            &  A         & -13.35       &  H-G99    &  7.46 (7.39 -- 7.53)     & H-G99       &{\it 7.76}           &{\bf 7.75}           \\ 
            &            &  B         &              &  H-G99    &  7.55 (7.50 -- 7.60)     & H-G99       &{\it 7.74}           &                     \\ 
DDO 190     &            &            & -15.10       &  H-G99    &  7.94 (7.88 -- 8.00)     & H-G99       &{\it 7.74}           &{\bf 7.74}           \\  \hline
\end{tabular}
\end{center}


{\it 1} -- in units of 12 + log(O/H)

{\it 2} -- $M_{pg}$ was taken as $M_{B}$

* -- oxygen abundance was derived not through the $T_{e}$ -- method

\vspace{0.05cm}

{\it List of references to} $M_{B}$:

H-G99 -- Hidalgo-G\'amez 1999;
H-GO -- Hidalgo-G\'amez \& Olofsson 1998;
KD   -- Kinman and Davidson 1981;
M    -- Mateo 1998;
MH94 -- Miller and Hodge 1994;
RM   -- Richer and McCall 1995;
SKH  -- Skillman, Kennicutt and Hodge 1989;
STKGT  --  Skillman, Terlevich, Kennicutt, Garnett, Terlevich, 1994; 
vZ     --  van Zee, 2000 

\vspace{0.05cm}

{\it List of references to oxygen abundances}:

G-RRZ  --  Gonz\'alez-Riestra, Rego, Zamorano, 1988; 
H-G99  --  Hidalgo-G\'amez 1999; 
HM     --  Hodge, Miller, 1995; 
KD     --  Kinman, Davidson, 1981; 
MAM    --  Moles, Aparacio, Masegosa, 1990; 
MH     --  Miller, Hodge, 1996; 
SKH    --  Skillman, Kennicutt, Hodge, 1989; 
STKGT  --  Skillman, Terlevich, Kennicutt, Garnett, Terlevich, 1994; 
STM    --  Skillman, Terlevich, Melnick, 1989; 
TBDS   --  Tully, Boesgaard, Dyck, Schempp, 1981; 
vZ     --  van Zee, 2000 
\end{table*}

The data used in this study consists of published absolute magnitudes of 
irregular galaxies in B band and the intensities of $[OII] \lambda \lambda 3727, 
3729$ and $[OIII] \lambda \lambda 4959, 5007$ emission lines. Our sample 
includes 34 data points in 20 irregular galaxies for which we have collected 
the relevant observational data, listed with references in Table \ref{table:OH}.
The commonly used name(s) of the galaxy are given in columns 1 and 2 and the 
label of the H\,{\sc ii} region or spectrum is given in column 3. The absolute 
blue magnitude $M_{B}$ is listed in column 4 (references in column 5). The 
original oxygen abundance is given in column 6 (references in column 7). If the 
error range of the oxygen abundance determination was indicated by the author(s), 
this range is given in parentheses (column 6). If a method other than the 
$T_{e}$ -- method has been used for the oxygen abundance determination, this 
oxygen abundance is labeled with an asterisk. 

\begin{figure}
\resizebox{1.00\hsize}{!}{\includegraphics[angle=0]{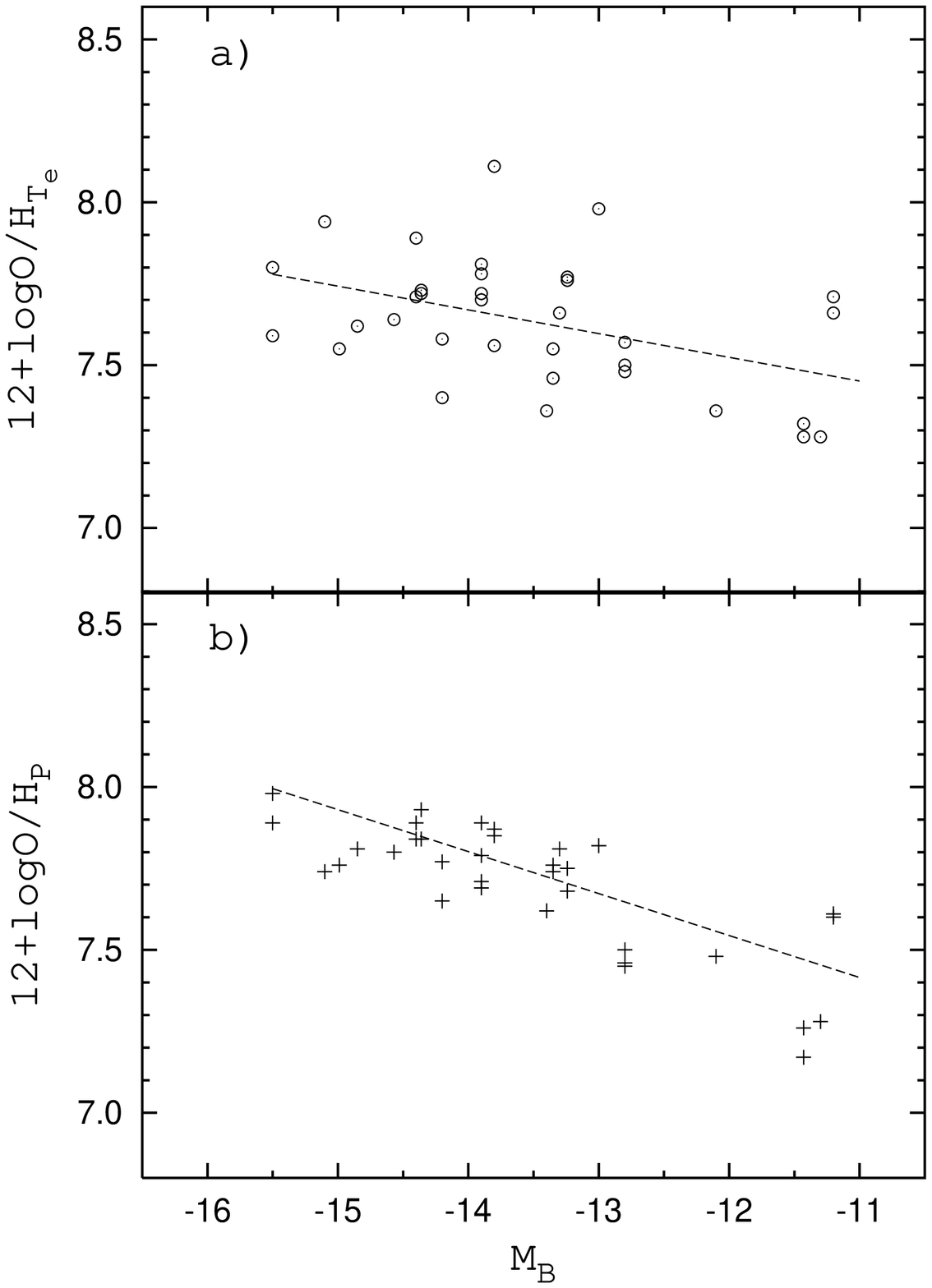}}
\caption{The metallicity -- luminosity diagram for dwarf irregular galaxies. 
The oxygen abundances for individual H\,{\sc ii} regions are shown. {\bf a)} 
The L-Z diagram based on the oxygen abundances derived with the $T_{e}$  -- 
method or the empirical method (see Table 1). The line is the best fit.
{\bf b)} The L-Z diagram based on the oxygen abundances determined here 
with the P -- method. The line is the best fit. }
\label{figure:z-l}
\end{figure}

The oxygen abundances in H\,{\sc ii} regions have been calculated with the P --
method using the published intensities (references in column 7) of 
$[OII] \lambda \lambda 3727, 3729$ and $[OIII] \lambda \lambda 4959, 5007$ 
emission lines. The derived oxygen abundances for individual H\,{\sc ii} regions 
are given in column 8 of Table 1. The metallicity -- luminosity diagram for 
dwarf irregular galaxies based on the oxygen abundances derived here through 
the P -- method is presented in Fig.\ref{figure:z-l} (panel {\it b}). The 
line is the best fit. For comparison, the metallicity -- luminosity diagram for 
dwarf irregular galaxies based on the oxygen abundances derived through the 
$T_{e}$ -- method or empirical method (data from the column 6 of Table 1) is 
also presented in Fig.\ref{figure:z-l} (panel {\it a}). The line is the 
best fit. 

Inspection of Fig.\ref{figure:z-l} shows that the metallicity -- luminosity
diagram for dwarf irregular galaxies based on the oxygen abundances derived 
with the P -- method shows relatively small scatter. It confirms the above 
conclusion that the large scatter in the low-luminosity end of the metallicity -- 
luminosity relation for irregular galaxies obtained by Hidalgo-G\'amez \& Olofsson 
(1998) is explained mainly by the uncertainty in oxygen abundance determination 
with $T_{e}$ -- method.

The uncertainty in the integrated B-band absolute magnitudes $M_{B}$ for 
irregular galaxies can also contribute to the dispersion in the Z -- L diagram.
The uncertainty in the distance determinations is the major source of
uncertainty in the determination of the integrated B-band absolute magnitudes
$M_{B}$ for irregular galaxies. Therefore accurate distances to irregular
galaxies are necessary in the construction of the Z -- L diagram.
Most galaxies considered here are members of the Local Group. The compilation
of the recent information on distances for all the dwarf members and candidates 
of the Local Group is given by Mateo (1998). Nearly all of the members of the 
Local Group  have reasonable distance determinations based on one or more 
high-precision distance indicator, including Cepheid variables. There are
exceptions. In the case of Leo A there is a large discrepancy in the determined 
distances; Hoessel et al. (1994) found a value of 2.2 Mpc on the basis of
Cepheid variables. More recently, Tolstoy et al. (1998) (based on the position 
of the red clump, the helium-burning blue loops, and the tip of the red giant
branch) obtained the value of 690 $\pm$ 60 kpc. The short distance resolves
the problems which appear for Leo A in the case of long distance (see discussion
in Mateo (1998) and references therein). The short distance makes the position 
of Leo A much closer to the metallicity -- luminosity relationship 
(see Fig.\ref{figure:hidalgo}). 

\begin{figure}
\resizebox{1.00\hsize}{!}{\includegraphics[angle=0]{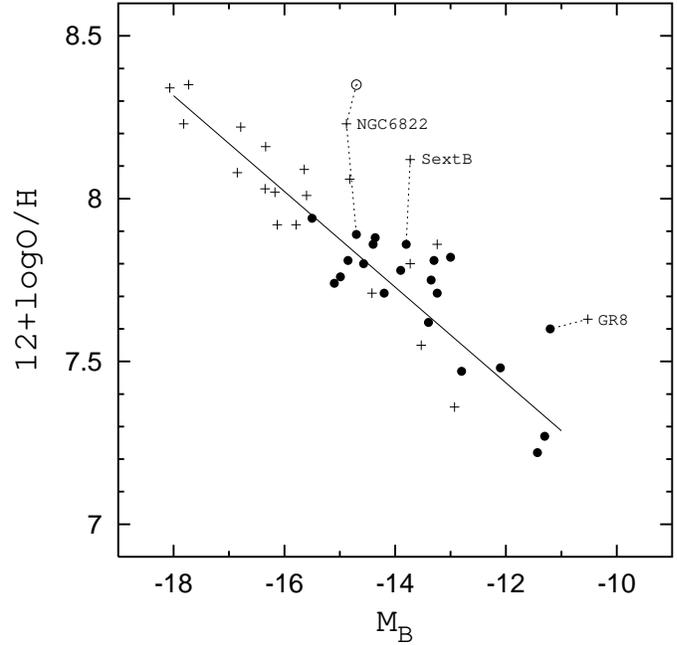}}
\caption{The metallicity -- luminosity diagram for dwarf irregular galaxies. The 
metallicity -- luminosity diagram of Richer \& McCall (1995) is represented by 
pluses. The metallicity -- luminosity diagram based on the oxygen abundances 
derived with the P -- method is shown by filled circles. The positions of 
Sextans B, GR8, and NGC6822 with our data and their positions with Richer and 
McCall's data are connected by dashed lines. The open circle is the position 
of NGC6822 with oxygen abundance from Eq.4.}
\label{figure:mccall}
\end{figure}

The comparison of our Z-L diagram with that of Richer \& McCall (1995) is given 
in Fig.\ref{figure:mccall}. The pluses are data of Richer \& McCall (1995), 
the filled circles are the data for irregular galaxies from the present sample. 
The oxygen abundances O/H$_{P}$ used in the construction of the Z-L diagram in 
Fig.\ref{figure:mccall} are the values obtained by averaging all the available 
determinations for a given galaxy (column 9 in Table 1).  The solid line is the 
metallicity -- luminosity relationship obtained by Richer \& McCall (1995) for 
the luminous ($M_{B}$ $<$ --15) irregular galaxies. As can be seen in 
Fig.\ref{figure:mccall}, the metallicity -- luminosity relation for 
low-luminosity dwarf irregular galaxies derived here is in good agreement with 
the metallicity -- luminosity relation obtained by Richer \& McCall (1995). 
However, the positions of some galaxies in our metallicity -- luminosity 
diagram and in the diagram of Richer \& McCall are appreciable different. 

According to the data of Richer \& McCall the positions of three galaxies
(Sextans B, GR8, and NGC6822) in the Z -- L diagram have large deviations 
from the metallicity -- luminosity relationship (Fig.\ref{figure:mccall}). 
The positions of Sextans B and GR8 in the Z -- L diagram with our data show 
significantly smaller deviations from the metallicity -- luminosity relationship 
than their positions  with Richer and McCall's data. (The positions of Sextans B, 
GR8, and NGC6822 in the Z -- L diagram with our data and positions with Richer 
and McCall's data are connected with dashed lines in Fig.\ref{figure:mccall}). 
There are two determinations of oxygen abundance in Sextans B 
(Table \ref{table:OH}) which result in different values of oxygen abundance:
12+logO/H=7.56 according to Skillman et al. (1989a) and 12+logO/H=8.11 according 
to Moles et al. (1990). The value of oxygen abundances from Moles et al. (1990) 
has been adopted by Richer \& McCall (1995). The oxygen abundances derived
through the P -- method with the line intensities measurements from 
Skillman et al. (1989a) and from Moles et al. (1990) are in good argeement:
12+logO/H=7.85 for the Skillman et al's data and 12+logO/H=7.87 for the 
Moles et al's data (Table \ref{table:OH}). This value of oxygen abundance
in the Sextans B is in agreement with oxygen abundance corresponding
to its luminosity according to the metallicity -- luminosity relationship.
A large deviation of the position of GR8 in the Richer and McCall Z -- L diagram 
from the metallicity -- luminosity relationship is caused mainly by the 
uncertainty in the value of the luminosity, Fig.\ref{figure:mccall}. The case of 
NGC6822 will be considered in the next subsection.

Thus, the metallicity -- luminosity diagram for low-luminosity dwarf irregular 
galaxies constructed here on the base of oxygen abundances derived through the
P -- method is in good agreement with the metallicity -- luminosity relation 
obtained by Richer \& McCall (1995). It should be particularly emphasized that 
Richer \& McCall have determined their metallicity -- luminosity relation
using only the best available data for dwarf irregulars. In contrast, we
have constructed the metallicity -- luminosity diagram using all the available 
data for dwarf irregulars. As a consequance of this, the number of points at the
low-luminosity end of the metallicity -- luminosity relation increases by
about a factor of two. As this takes place, the scatter at low luminosities in
our diagram is not in excess of that in Richer and McCall's diagram, although 
the dispersion of oxygen abundances around the metallicity -- luminosity 
relationship seems to increase with decreasing of galaxy luminosity, as was 
found by Richer \& McCall (1995). 

\subsection{The NGC6822}

The position of  NGC6822 in the Z -- L diagram shows a large deviation from the 
metallicity -- luminosity relationship (Fig.\ref{figure:mccall}). The oxygen 
abundance determinations both in individual stars and in H\,{\sc ii} regions in 
the NGC6822 are now available. The mean stellar oxygen abundance derived from 
high resolution spectra of two stars is 12+logO/H$_{star}$ = 8.36$\pm$0.19 
(Venn et al. 2001). Twelve determinations of oxygen abundances in H\,{\sc ii} 
regions of the NGC6822 (Lequeux et al. 1979, Pagel et al. 1980, Skillman et al. 
1989b, Hidalgo-G\'amez 1999) result in a mean value of gas oxygen abundance 
12+logO/H = 8.23 (Table \ref{table:ngc6822}). If this value of gas oxygen 
abundance in the NGC6822 is correct, then the position of NGC6822 in the 
metallicity -- luminosity diagram deviates considerably from the metallicity -- 
luminosity relationship, i.e. this value of oxygen abundance in the NGC6822 is
significantly (by around 0.3dex) higher than the oxygen abundance corresponding
to its luminosity, Fig.\ref{figure:mccall}. 

\begin{table}
\caption[]{\label{table:ngc6822}
Oxygen abundances in H\,{\sc ii} regions of the dwarf irregular galaxy NGC6822.
The name of the H\,{\sc ii} region is reported in column 1. The original oxygen abundance 
is reported in column 2 (reference in column 3). The oxygen abundance derived 
through the T$_{e}$ -- method is labeled "Te". The oxygen abundance 
derived here through the P -- method is listed in column 4. 
}
\begin{center}
\begin{tabular}{llcc} \hline \hline
           &                    &            &                  \\  
H\,{\sc ii} region & 12+logO/H          & reference  &12+log(O/H)$_{P}$ \\  
           &                    &            &                  \\   \hline
Ho 11      & 8.24$\pm$0.20      &  Petal     &    8.25          \\   
Ho 12      & 8.23$\pm$0.20      &  Petal     &    8.32          \\   
Hu X       & 8.21$\pm$0.15 (Te) &  Petal     &    8.44          \\   
Hu V       & 8.20$\pm$0.09 (Te) &  Petal     &    8.40          \\   
Ho 14      & 8.27$\pm$0.20      &  Petal     &   (8.01)         \\   
Ho 13      & 8.50$\pm$0.20      &  Petal     &    8.32          \\   
Ho 15      & 8.11$\pm$0.12 (Te) &  Petal     &    8.31          \\   
Hu X       & 8.27$\pm$0.06 (Te) &  Letal     &    8.34          \\   
Hu V       & 8.20$\pm$0.06 (Te) &  Letal     &    8.38          \\   
Hu V       & 8.20$\pm$0.12      &   STM      &    8.37          \\   
Hu V       & 8.13$\pm$0.02 (Te) &   H-G      &    8.38          \\   
Hu X       & 8.09$\pm$0.06 (Te) &   H-G      &    8.34          \\   
mean       & 8.23               &            &    8.35          \\     \hline  \hline 
\end{tabular}
\end{center}

\vspace{0.05cm}

{\it List of references}:

H-G    --  Hidalgo-G\'amez 1999;
Letal  --  Lequeux et al. 1979;
Petal  --  Pagel et al. 1980; 
STM    --  Skillman, Terlevich, and Melnick 1989
\end{table}

If H\,{\sc ii} regions in NGC6822 lie on the lower branch of the R$_{23}$ -- O/H 
diagram then,  Eq.(\ref{equation:oh-p}) can be used for the oxygen abundance 
determination in H\,{\sc ii} regions of NGC6822. The mean value of oxygen 
abundances obtained via Eq.(\ref{equation:oh-p}) is around 7.90. In this case 
the position of the NGC6822 in the Z -- L diagram is close to the metallicity -- 
luminosity relationship, Fig.\ref{figure:mccall},  but this value of oxygen 
abundance is in conflict with a stellar oxygen abundance 12+logO/H$_{star}$  
and with the oxygen abundances of O/H$_{T_{e}}$ derived through the T$_{e}$ -- 
method (Table \ref{table:ngc6822}). 

\begin{figure}
\resizebox{1.00\hsize}{!}{\includegraphics[angle=0]{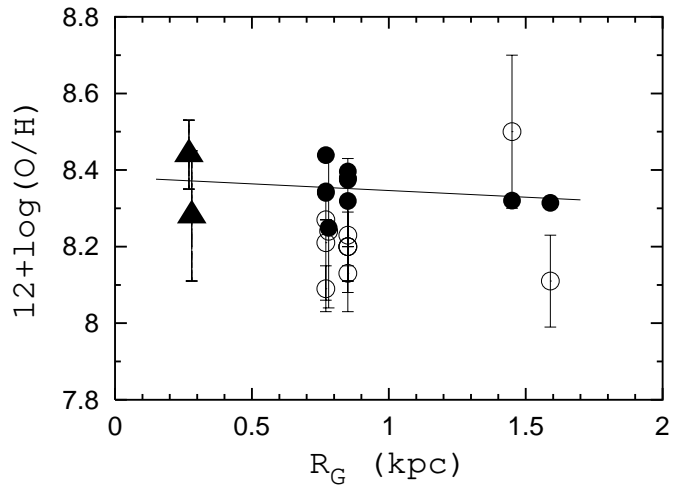}}
\caption{
The radial abundance gradient within NGC6822. The original oxygen abundances 
of H\,{\sc ii} regions (column 2 of Table 2) are represented by open circles 
with error bars. The stellar data (with error bars) are shown by filled 
triangles. Our oxygen abundances O/H$_{P}$ for H\,{\sc ii} regions
(column 4 of Table 2) are shown by filled circles. The line is the best fit 
to our data for H\,{\sc ii} regions  and to the original data for stars. }
\label{figure:grad6822}
\end{figure}

The stellar oxygen abundances and oxygen abundances derived through the T$_{e}$ 
-- method suggest that the H\,{\sc ii} regions in NGC6822 lie on the upper 
branch of the R$_{23}$ -- O/H diagram. Then, the corresponding equations from 
Paper II 
\begin{equation}
12+log(O/H)_{P} = \frac{R_{23} + 54.2  + 59.45 P + 7.31 P^{2}}
                       {6.07  + 6.71 P + 0.371 P^{2} + 0.243 R_{23}}
\label{equation:OHR23f}
\end{equation}
should be used for the oxygen abundance determination in H\,{\sc ii} regions 
of NGC6822. In this case the mean oxygen abundance for 11 H\,{\sc ii} regions in 
NGC6822 is 12+logO/H$_{P}$ = 8.35, Table \ref{table:ngc6822} (the value of 
O/H$_{P}$ obtained in the H\,{\sc ii} region Ho 14 from the sample of Pagel 
et al.(1980) is in conflict with the suggestion that this H\,{\sc ii} region 
belongs to the upper branch of the R$_{23}$ -- O/H diagram, and this H\,{\sc ii} 
region is excluded from the consideration).  The mean value of O/H$_{P}$ is in 
agreement with the mean value of the stellar oxygen abundance.  But in this case 
the position of the NGC6822 in the Z  -- L diagram deviates considerably from the 
metallicity -- luminosity relationship (open circle in Fig.\ref{figure:mccall}). 

The values of the distance modulus of NGC6822 determined with different 
distance indicators (Cepheids, the tip of the red giant branch) are in good 
agreement (Lee et al. 1993, Gallart et al. 1996, Ferrarese et al. 2000); the
uncertainty seems to be not in excess of $\pm$0.15. Therefore the deviation of 
NGC6822 from the metallicity -- luminosity relation cannot be caused by the 
uncertainty in the luminosity determination. Thus, the large deviation of 
NGC6822 from the metallicity -- luminosity relationship seems to be real, 
although the actual value of the deviation is not clear.

In Fig.\ref{figure:grad6822} we show our oxygen abundances O/H$_{P}$
for H\,{\sc ii} regions from Table \ref{table:ngc6822}, as a function of galactocentric 
distance, together with the original oxygen abundances in H\,{\sc ii} regions and
with data for stars. The original oxygen abundances are presented 
by open circles with error bars. The stellar data (with error bars) are 
shown by filled triangles. Our data are shown with filled circles. 
The galactocentric distances for H\,{\sc ii} regions and stars were taken from 
Venn et al. (2001). The line is the best fit to our data for H\,{\sc ii} regions 
and to the original data for stars. From examination of the O/H$_{P}$ in the 
H\,{\sc ii} regions together with the stellar data, it is evident that there is no 
significant radial abundance gradient within the NGC6822. The slope of the 
formal best fit is about --0.035 dex/kpc. The data are also consistent 
with no gradient at all. Thus, our data confirm the conclusion of Pagel et al. 
(1980) that if there is an actual radial abundance gradient within NGC6822, 
the slope of the gradient is small and the trend is entirely masked by the errors.

\section{Discussion and conclusions}

Thus, the metallicity -- luminosity relation for irregular galaxies extends into 
the region of low luminosities up to M$_B$ $\sim$ --12. It is widely suggested 
that the metallicity -- luminosity relation for irregular galaxies is caused by 
galactic winds of different efficiencies. In other words, the metallicity -- 
luminosity relation represents the ability of a given galaxy to keep the products 
of its own evolution rather than their ability to produce metals (Larson 1974). 
On the other hand, it has been found that the astration level is higher in 
massive irregular galaxies than in dwarf ones (Lequeux et al. 1979, Vigroux 
et al. 1987). Then, it is suggested that the systematic increase of the 
astration level with luminosity can aslo play a role in the origin of the 
metallicity -- luminosity relationship for irregular galaxies.

The values of the gas mass fraction $\mu$ and oxygen abundance deficiency (which
is a good indicator of the efficiency of mass exchange between a galaxy and its 
environment) have been derived for a number of late-type spiral (Pilyugin \&
Ferrini 1998) and irregular (Pilyugin \& Ferrini 2000a) galaxies. Using these
data, the roles of two hypothesed mechanisms as causes of the 
metallicity -- luminosity correlation among late-type spiral and irregular 
galaxies have been examined in our recent study (Pilyugin \& Ferrini 2000b).
It was found that both the increase in astration level and the decreasing
efficiency of heavy element loss with increasing luminosity, make comparable 
contributions to the metallicity -- luminosity correlation. The fact that the 
metallicity -- luminosity relation for irregular galaxies extends up to low 
luminosity  can be considered as evidence that the tendency for the decrease 
in astration level and the increasing efficiency of heavy element loss with 
decreasing luminosity remains at the low-luminosity end.

A prominent feature of the metallicity -- luminosity relation for irregular 
galaxies is the increased scatter at low luminosities in comparison to that at 
high luminosities.  Part of this scatter is undoubtedly due to uncertainties in 
the oxygen abundances and luminosities. But another part is likely to be real. 
Since there is no apparent reason to suggest that the uncertainties in oxygen 
abundances and/or in luminosities increase systematically with decreasing 
galaxy luminosity, the increase of the dispersion of oxygen abundances 
around the metallicity -- luminosity relationship with decreasing galaxy 
luminosity seems to be real. The increase in the scatter of oxygen abundances 
with decreasing of luminosity can be explained by the fluctuations in values 
of the gas mass fraction among irregular galaxies of a given luminosity. 
There is a correlation between the gas mass fraction $\mu$ and the galaxy's
luminosity, decreasing from $\mu$ $\sim$ 0.8 at M$_B$ = -12 or $\log$L$_{B}$ = 7
to $\mu$ $\sim$ 0.4 at M$_B$ = -18 or $\log$L$_{B}$ = 9.4 (Pilyugin \& 
Ferrini 2000a). According to the simple model for the chemical evolution
of galaxies, the relation between oxygen abundance and gas mass fraction is given 
by a logarithmic relationship, $Z_{O}$ $\sim$ $\ln$(1/$\mu$), then the
O/H -- $\mu$ curve is significantly steeper at large than at small $\mu$.
Therefore, the equal fluctuations in values of gas mass fraction, say 
$\Delta \mu$ = 0.1, result in an appreciably larger dispersion 
of oxygen abundances among low-luminosity irregular galaxies with high values 
of gas mass fraction (the change of $\mu$ from 0.9 to 0.8 results in 
$\Delta \log$O/H $\sim$ 0.33) than among luminous irregular galaxies with low 
values of the gas mass fraction (the change of $\mu$ from 0.5 to 0.4 results in 
$\Delta \log$O/H $\sim$ 0.12). Thus, the increase in the scatter of oxygen 
abundances with decreasing of luminosity is not surprising. 

{\it In summary:}

The low-luminosity dwarf irregular galaxies are considered.
The oxygen abundances in H\,{\sc ii} regions of dwarf irregular galaxies  were 
recalculated from published spectra through the recently suggested P --
method. It has been found  that the metallicity of low-luminosity dwarf 
irregular galaxies, with a few exceptions, correlates well with galaxy 
luminosity. The dispersion of oxygen abundances around the luminosity -- 
metallicity relationship increases with decreasing galaxy luminosity,
as was found by Richer \& McCall (1995). The lack of relationship between the 
oxygen abundance and the absolute magnitude in the blue band for irregular
galaxies obtained by Hidalgo-G\'amez \& Olofsson (1998) can be explained by the
large uncertainties in the oxygen abundances derived through the $T_{e}$ --
method, that in turn can be explained by the large uncertainties in the 
measurements of the strengths of the weak oxygen line $[OIII] \lambda 4363$ 
used in the $T_{e}$ -- method.

\begin{acknowledgements}
It is a pleasure to thank the referee, Dr. N.~Arimoto, for his comments on this 
work. This study was partly supported by the NATO grant PST.CLG.976036 and 
the Joint Research Project between Eastern Europe and Switzerland (SCOPE) 
No. 7UKPJ62178.
\end{acknowledgements}

\end{document}